# Simulation of bifurcated stent grafts to treat abdominal aortic aneurysms (AAA)

J. Egger[ab], S. Großkopf[b], B. Freisleben[a]

[a]Philipps-University of Marburg, Dept. of Mathematics and Computer Science, Germany;
[b]Siemens Computed Tomography, Forchheim, Germany

## ABSTRACT

In this paper a method is introduced, to visualize bifurcated stent grafts in CT-Data. The aim is to improve therapy planning for minimal invasive treatment of abdominal aortic aneurysms (AAA). Due to precise measurement of the abdominal aortic aneurysm and exact simulation of the bifurcated stent graft, physicians are supported in choosing a suitable stent prior to an intervention. The presented method can be used to measure the dimensions of the abdominal aortic aneurysm as well as simulate a bifurcated stent graft. Both of these procedures are based on a preceding segmentation and skeletonization of the aortic, right and left iliac. Using these centerlines (aortic, right and left iliac) a bifurcated initial stent is constructed. Through the implementation of an ACM method the initial stent is fit iteratively to the vessel walls – due to the influence of external forces (distance- as well as balloonforce). Following the fitting process, the crucial values for choosing a bifurcated stent graft are measured, e.g. aortic diameter, right and left common iliac diameter, minimum diameter of distal neck. The selected stent is then simulated to the CT-Data – starting with the initial stent. It hereby becomes apparent if the dimensions of the bifurcated stent graft are exact, i.e. the fitting to the arteries was done properly and no ostium was covered.

## 1. INTRODUCTION

An aneurysm is – in contrast to an arterial stenosis which is an abnormal narrowing – a dilation of a blood vessel. In case of abdominal aortic aneurysms, reaching a critical diameter above 55 millimeters the risk of rupture increases. Once ruptured it leads to internal bleedings which causes death in the most cases – even if the affected artery is operated immediately. If known an aneurysm could be treated before the rupture occurs. This is done by deploying a stent graft, which splints and eliminates the aneurysm. Fitting correctly inside the affected artery, the stent graft overtakes the blood flow in the aneurysm area. Optimally the stent graft excludes the aneurysm from the blood flow and the aneurysm sack degenerates.

There are two possibilities to treat an aneurysm with a graft before a rupture occurs. One option is the open surgery. Thereby the patient's body is opened at the location of the aneurysm and then the graft is sewed into vascular tissue. But this form of therapy is very stressful on the patient and not eligible for everyone, e.g. risk patients. An alternative therapy is the endovascular surgery. Insertion of the stent graft is herby done by catheter technique – mostly thru a small cut in the femoral – and requires a precise determination of the aneurysm dimension for planning the operation.

Randomized multicenter studies of abdominal aortic aneurysms (EVAR, DREAM) have shown that the mortality rate of endovascular surgery is lower within the first 24 months after surgery[1, 2]. After 2 years, the mortality rate is similar to open surgery, e.g. caused by occurring endoleaks (Fig. 1). Improved support of patient selection, planning and follow-up, based on improved imaging and image processing, will hopefully reduce the increasing mortality rate. Therefore we want to measure the affected artery and simulate and visualize stents preoperatively in the CT-Data.

---

Send correspondence to Jan Egger: jan.egger.ext@siemens.com

In previous works[3, 4] different methods were introduced for simulation and visualization of stent grafts which are not bifurcated. In this paper we focus on the more complex bifurcated stent grafts. This type of stent graft is used for abdominal aortic aneurysms (AAA) which can not be treated with simple tube stent grafts because the aneurysm ranges from the aorta to far into the iliac arteries.

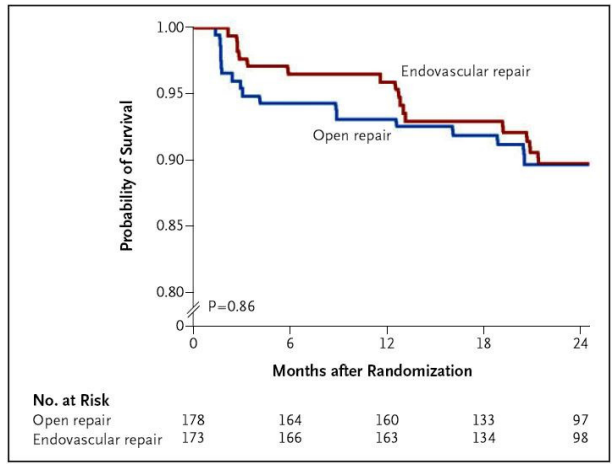

Fig. 1. Mortality rate after open- and endovascular surgery[2]

## 2. METHODS

For measurement of the abdominal aortic aneurysm and bifurcated stent graft simulation, the affected artery is segmented with a region growing method that starts at a user-defined seed point. Then, a skeleton algorithm determines the three vessel centerlines (aorta, right and left iliac) by iterative erosion of the segmentation mask[5]. Starting with these centerlines an initial bifurcated stent graft is constructed with a given radius (Fig. 2).

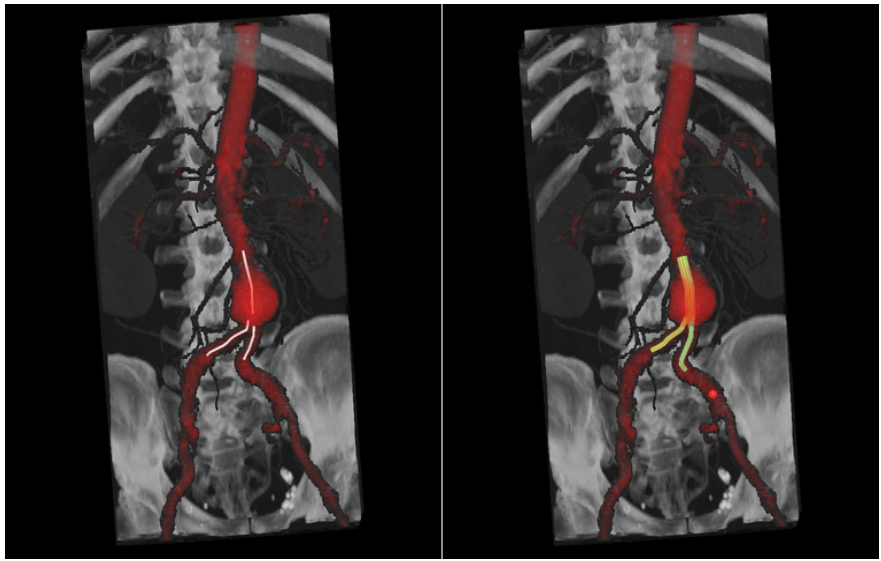

Fig. 2. Left: Vessel centerlines (right iliac, aorta and left iliac), right: Initial stent

After this pure geometrical bifurcated stent graft construction as it was described in[3], the bifurcated stent graft is deformed by using an active contours (ACM) method. These methods were used, to:

- Simulate the stent behavior, and
- Measure critical diameters and lengths to select an appropriate stent

### 2.1 Simulate the stent behavior

The ACM method bases on the technique first published by Kass et al.[6, 7] and was extended to a two-dimensional surface in three-dimensional space[8, 9] that minimizes the following energy functional:

$$E = \int_{t=0}^{1} \int_{s=0}^{1} E_{int}(v(s,t)) + E_{ext}(v(s,t)) ds dt \qquad (1)$$

The physical attributes of the virtual bifurcated stent graft are simulated by internal forces in horizontal, vertical and diagonal directions (2) and different external forces (3):

$$E_{int} = w_1 \frac{\partial v(s,t)}{\partial s} + w_2 \frac{\partial v(s,t)}{\partial t} + w_3 \frac{\partial^2 v(s,t)}{\partial s^2} + w_4 \frac{\partial^2 v(s,t)}{\partial t^2} + w_5 \frac{\partial^2 v(s,t)}{\partial s \partial t} \qquad (2)$$

$$E_{ext} = w_{vesselWall} \cdot D(x,y,z) + w_{balloon} \cdot F_{balloon} \qquad (3)$$

To achieve the aim of minimizing the energy functional of equation (2, 3), the belonging Euler-Lagrange equations have to be solved:

$$-\frac{\partial}{\partial s}\left(w_1 \frac{\partial v}{\partial s}\right) - \frac{\partial}{\partial t}\left(w_2 \frac{\partial v}{\partial t}\right) + \frac{\partial^2}{\partial s^2}\left(w_3 \frac{\partial^2 v}{\partial s^2}\right) + \frac{\partial^2}{\partial t^2}\left(w_4 \frac{\partial^2 v}{\partial t^2}\right) + 2\frac{\partial^2}{\partial s \partial t}\left(w_5 \frac{\partial^2 v}{\partial s \partial t}\right) = -\nabla E_{ext} \qquad (4)$$

One external force is called the balloon force $F_{balloon}$ which simulates the balloon that is used to expand the bifurcated stent graft:

$$F_{balloon} = \begin{cases} F_{pressure}(R-r) & if \|r\| < \|R\| \\ 0 & otherwise \end{cases} \qquad (5)$$

$R$ is the radius which restricts the virtual stent to expand further than the system of stent and balloon can do.

Another external force $F_{vesselWall} = \nabla D(x,y,z)$ is derived from the Euclidian DTF of the segmented artery. It simulates the resistance of the vessel wall against expansion. Voxels inside the artery are initialized with zero whereas voxels outside are assigned with the minimum Euclidean distance from the vessel wall.

The influence of the internal forces is simulated by setting up one stiffness matrix for the complete bifurcated stent. We are able to create it by linking the three segments with a connection element (see fig. 3, right). This connection element defines for every segment boundary local predecessor vertices and thus enables the numerical approximation of derivatives up to an order of four.

### 2.2 Measure some critical diameters and lengths to find an appropriate stent

For measuring the dimensions of the AAA, the external forces press respectively pull on the bifurcated stent graft in the direction toward the vessel walls. By minimizing $w_1 ... w_5$ while maximizing $R$, the internal forces loose their influence on the simulation. Thus the bifurcated stent graft expands to completely fit to the vessel walls. From this geometric model of the AAA the relevant sizes for choosing a bifurcated stent graft can be measured easily (Fig. 3, left).

In detail we measure the following diameters, which can be determined from CT images only:

- Aortic diameter at proximal implantation site (a)
- Aortic diameter – 15 mm inferior to proximal implantation site (b)
- Maximum aneurysm diameter (c)
- Minimum diameter of distal neck (d)
- Right common iliac diameter (e)
- Left common iliac diameter (f)

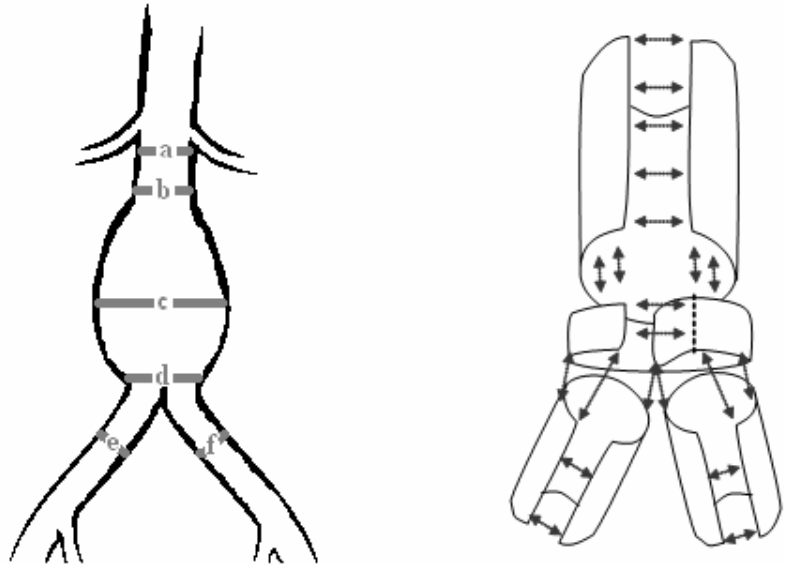

Fig. 3. Measurements taken from the AAA (left) and impact of the internal forces (right)

## 3. RESULTS

The method presented in this paper was implemented in C++ in the MeVisLab environment. Results are demonstrated for CTA with variations in anatomy and location of the pathology. For testing we used two kinds of CT data. One dataset came from real CT scans acquired during clinical routine. The other dataset consists of artificially generated CT abdominal aortic aneurysm data to verify the implementation.

The ACM method for simulating a bifurcated stent grafts provided good results. The material properties of the stent grafts were simulated suitably and the fit to the vessel wall was realistic (Fig. 4 and 5). Using a CT dataset with 512*512*385 voxels and a stent graft consisting of 624 surface vertices, the calculation of the inverse stiffness matrix took in our implementation about 80 seconds on an Intel Xeon CPU, 3 GHz, 3 GB RAM, Windows XP Professional 2002. An iterative expansion step within the ACM method took less than one second.

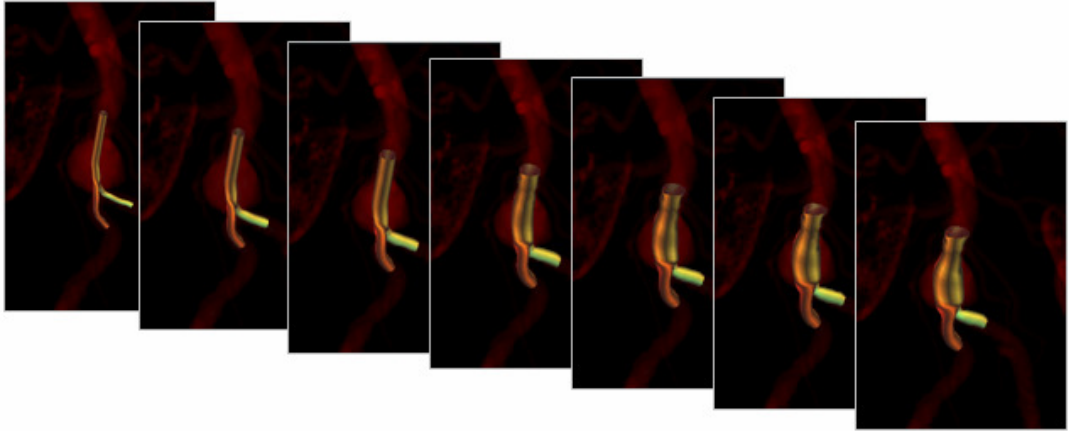

Fig. 4. Simulation of a bifurcated stent graft (left: initial stent, right: stent after 50 iterations)

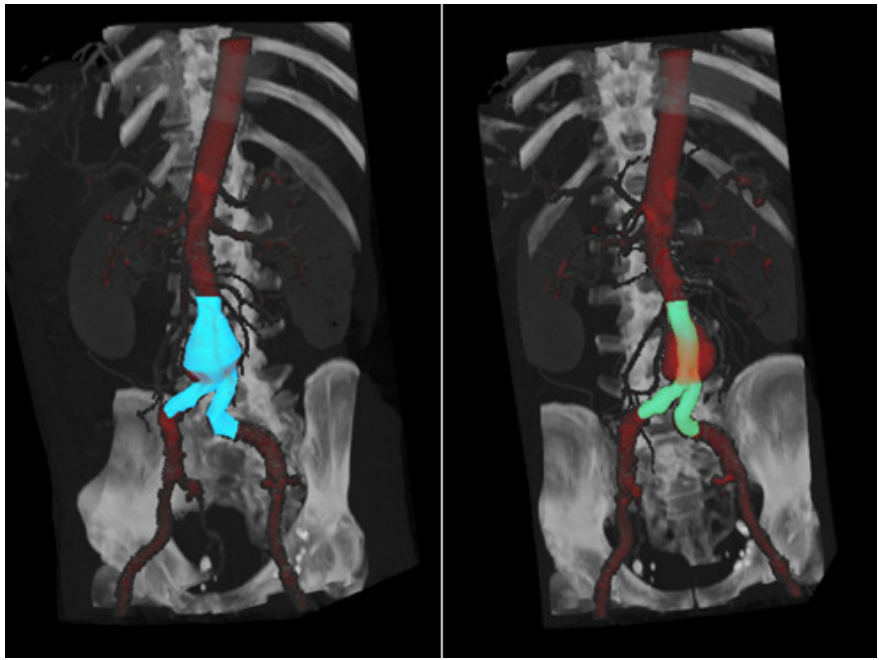

Fig. 5. Results of the segmentation of an AAA (left) and the simulation of a bifurcated stent graft (right)

## 4. CONCLUSIONS

We segmented abdominal aortic aneurysm and measured the dimensions of these aneurysms. With these measured values we constructed and visualized bifurcated stent grafts in the CT-Data. The method provides realistic results for the simulation of bifurcated stent grafts. Based on this simulation, physicians are supported in choosing a bifurcated stent graft before an intervention. This is very important because a bifurcated stent graft which has not the exact dimensions could shift or cover an artery branch.